# Shedding new light on the mystery of wetting on soft solids

Su Ji Park[1], Byung Mook Weon[2], Ji San Lee[1], Junho Lee[1], Jinkyung Kim[1] & Jung Ho Je[1]*

[1]X-ray Imaging Center, Department of Materials Science and Engineering, Pohang University of Science and Technology, San 31, Hyoja-dong, Pohang 790-784, South Korea.

[1]School of Advanced Materials Science and Engineering, SKKU Advanced Institute of Nanotechnology (SAINT), Sungkyunkwan University, Suwon 440-746, South Korea.

One of the most questionable issues in wetting is the vertical force balance that is excluded in Young's law. On soft deformable solids, such as biotic materials and synthetic polymers, the vertical force of liquid leads to a microscopic protrusion of the contact line, i.e. a "wetting ridge". The wetting principle that determines the geometry of the ridge-tip is at the heart of the issues over the past half century. Here, we reveal a universal wetting principle by directly visualizing ridge-tips with high spatio-temporal resolution using x-ray microscopy. We find that the tip-geometry is asymmetric and bent, and invariant during ridge growth or by surface softness. This singular geometry is explained by linking the macroscopic and microscopic contact angles to Young's and Neumann's laws, respectively. Our dual-scale model would be applicable to a general framework in wetting and give new clues to various issues in cell-substrate interaction and elasto-capillary problems.

Many natural and synthetic materials are soft and deformable due to the nature as in rubbers or the structure like thin films[1–3] and fibers[4,5]. During recent decades, much attention has been paid to wetting behaviors of soft deformable solids[1–19] for various scientific issues involved in soft tissues[20–25] and polymer gels[26], as well as for practical applications such as inkjet printing[27] and microfluidic devices[28]. Wetting behaviors on solids have been explained by Young's law ($\cos\theta_L = (\gamma_{SV} - \gamma_{SL})/\gamma_{LV}$)[29], based on a force balance of interfacial tensions ($\gamma_{LV}$, $\gamma_{SV}$ and $\gamma_{SL}$ of liquid–vapor, solid–vapor and solid–liquid interfaces, respectively) that determines the equilibrium contact angles ($\theta_S$, $\theta_L$ and $\theta_V$ in solid, liquid and vapor, respectively) at the three-phase contact line.

On a soft solid, however, the vertical surface tension of liquid, which is excluded in Young's law, plays critical roles in the wetting behaviors, such as the wrinkling[1,2] or folding[3] of thin films and the bending of fibers[4,5]. The vertical force, highly concentrated at the contact line, induces a local microscopic structure on the soft surface, i.e. a "wetting ridge". This recondite feature makes questionable the validity of the classical law. Despite many theoretical[6–10] and experimental[6,11–18] studies over recent decades, uncertainty surrounding the microscopic geometry of a wetting ridge, particularly its "tip", still remains. Direct observation of the tip for precise measurements of the contact angles is hardly achievable by optical imaging due to limited resolutions[6,12–14]. Here, we present the first, direct and real-time visualization of ridge-tips with high spatial and temporal resolutions using x-ray microscopy and discuss a general framework of wetting on soft solids based on precise measurement of the geometry of the tips.

## Results

**Direct visualization of ridge-tips.** For this study, we used transmission x-ray microscopy (TXM), as illustrated in Fig. 1a, to provide direct visualization of a wetting ridge for a water (or 40% ethylene glycol aqua solution (EG 40%)) drop of $r \approx 1$ mm on a silicone gel (or polydimethylsiloxane (PDMS)) surface (Fig. 1b). All of the three interfaces, particularly including the liquid–vapor interface, at the tip of a wetting ridge were clearly visualized, as demonstrated in Figs 1c and d (white square of Fig. 1c), with a high spatial-resolution (~50 nm/pixel) without using any contrast agents. The interference bright and dark fringes at each interface, originating from the Zernike phase contrast[31,32], allowed us to clearly identify each interface[33,34]. We found that wetting ridges have actually asymmetric and bent cusps (Fig. 1d), instead of the known symmetric and triangular cusps measured using an indirect optical imaging method[6,13]. High-resolved images of the tips of cusps enabled us to accurately measure the microscopic ($\theta_S$, $\theta_L$ and $\theta_V$) as well as the macroscopic ($\theta$) contact angles (Fig. 1e; white square of Fig. 1d).

**Solid elasticity.** The direct visualization made it possible to accurately extract the surface profiles $u_{zx}$ (vertical surface displacement at $x$) of wetting ridges (see Supplementary Fig. S1), as demonstrated for two surfaces with $E$ (elasticity) = 3 and 16 kPa in Figs 2a and b. The two profiles (blue and pink circles in Fig. 2b) clearly show asymmetric and bent cusps that are significantly deviated from the best fits (dashed lines) of a linear elastic (LE) model[6,7] in of the region $w$ (ridge width) $\ll l_e$ (= $\gamma_S/E$, the elasto-capillary length). More interestingly, we found that the shapes of two cusps are nearly identical or self-similar (the inset of Fig. 2b), regardless of $E$ or the ridge height $u_{z0}$ ($\sim \gamma_{LV}\sin\theta /E$)[6,7,10,14]. Here, the LE model assumed $\theta = 90°$ and $\gamma_S = (\Upsilon_{SL} + \Upsilon_{SV})/2$, where $\Upsilon_{SL}$ and $\Upsilon_{SV}$ are the surface stresses at the SL and SV interfaces, respectively. The horizontal peak position of each surface profile was set as $x = 0$.

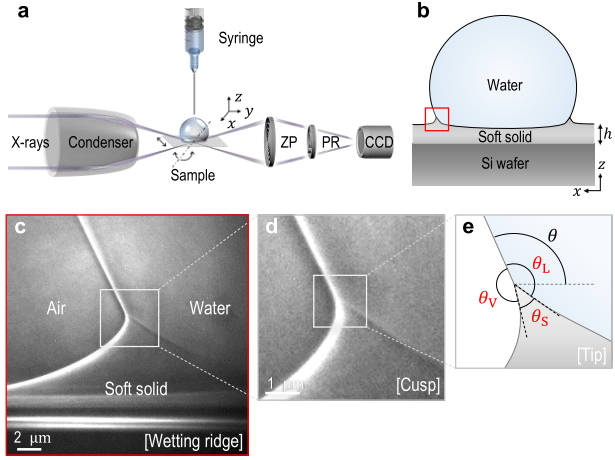
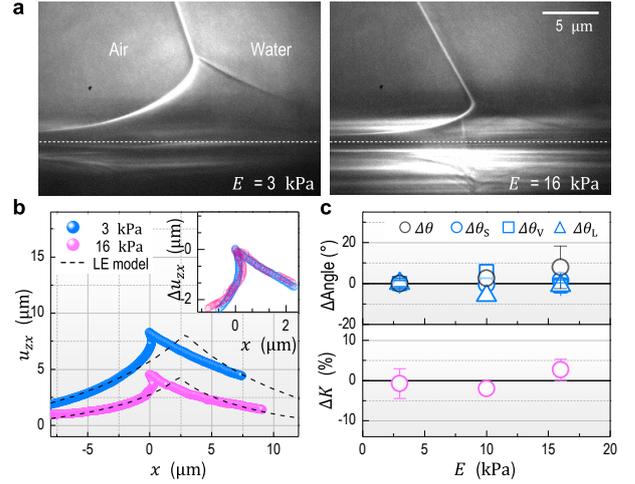

**Figure 1 | High resolution x-ray imaging for wetting ridge formation at three-phase contact line. a, b,** Schematic illustrations of (**a**) transmission x-ray microscopy (TXM) and (**b**) a sessile drop on a soft substrate. TXM consists of a capillary condenser, a motorized sample stage, a zone plate (ZP), a phase ring (PR) and a CCD camera. A wetting ridge (red square in **b**) is formed by the surface tension of a water drop at the contact line. **c, d,** Demonstration of a directly visualized wetting ridge (**c**) with an asymmetric and bent cusp (**d**; white square in **c**). **e,** Extraction of three interfaces from the ridge tip (white square in **d**), which enables us to measure the macroscopic ($\theta$) and the microscopic ($\theta_S$, $\theta_V$ and $\theta_L$) contact angles.

**Figure 2 | Effect of surface elasticity on wetting ridge formation. a,** Representative x-ray images of wetting ridges on a silicone gel ($E \approx 3$ kPa) and a PDMS film (16 kPa). **b,** The surface profiles, extracted from the images (**a**) using MATLAB and Image Pro-Plus 6.0 software (Supplementary Fig. S1), clearly show a strong $E$-dependence of the vertical displacement $u_{zx}$. The asymmetric and bent cusps are compared to the linear elastic (LE) model[6,7] (dashed lines). (Inset) The cusps are identically superimposed by $\Delta u_{zx} = u_{zx} - u_{z0}$ at $w << l_e$. **c,** (Top) The angular differences ($\Delta$Angle) are plotted with $E$, based on the average values for 3 kPa, for the macroscopic ($\theta$, dark gray circle) and the microscopic ($\theta_S$ (blue circle), $\theta_V$ (blue square) and $\theta_L$ (blue triangle)) angles measured at $w < \sim 0.4$ μm. The error bars are standard deviations from 5 sets of image data for 3 kPa, 1 set for 10 kPa and 3 sets for 16 kPa. (Bottom) $K$ is obtained with $\theta_S$ and $\gamma_{LV}$, i.e. $K = \sin\theta_S/\gamma_{LV}$. The little difference of $K$ ($\Delta K <$ 4%), resulted from that of $\Delta\theta_S$ ($< \pm 1.6°$), indicates invariant surface stresses $\Upsilon_{SL}$ and $\Upsilon_{SV}$ at the tips.

The macroscopic and the microscopic contact angles (Fig. 1e) were measured in a length of ~0.4 μm for various surface elasticities, as listed in Supplementary Table S1 and the angular dependencies on surface elasticity were plotted in Fig. 2c (top) based on the averaged values for $E \approx 3$ kPa. Interestingly, we found that the variation of each angle by elasticity is very small despite the strong elasticity-dependence of $u_{zx}$. In particular, the small difference in $\Delta\theta_S$ ($< \pm 1.6°$) leads to a nearly constant $K$ ($= \sin\theta_S/\gamma_{LV}$) with $E$ as in Fig. 2c (bottom, $\Delta K < 4\%$), directly indicating that the two surface stresses of $\Upsilon_{SL}$ and $\Upsilon_{SV}$ are invariant with $E$, as deduced from Neumann's relation ($K = \sin\theta_S/\gamma_{LV} = \sin\theta_V/\Upsilon_{SL} = \sin\theta_L/\Upsilon_{SV}$)[13]. Herein surface stresses[30] (specifically, $\Upsilon_{SV}$ and $\Upsilon_{SL}$), instead of surface energies ($\gamma_{SV}$ and $\gamma_{SL}$), are adopted as effective interfacial tensions for SV and SL interfaces[13], which also helps to overcome a paradox: a force balance at the tip but the violation of the Neumann triangle condition for $\gamma_{W(or\ EG\ 40\%)} > \gamma_{PDMS} + \gamma_{W(or\ EG\ 40\%)-PDMS}$ (see Supplementary Table S2).

Here a little large deviation in the macroscopic angle ($\theta$) presumably results from drop evaporation and surface heterogeneity. We found that the macroscopic angle ($\theta = 108.1 \pm 9.0°$) for the soft surfaces is consistent to that ($\theta = 106.6 \pm 2.3°$) for rigid surfaces ($E \approx 750$ kPa; see Methods Summary). This suggests that Young's law would hold regardless of $E$. We note that herein Young's law is valid since the surface stresses are equal to the surface energies for flat surfaces[30]. The Laplace pressure is negli-gible for large droplets as in our case, different for small (< 250 μm) droplets[13].

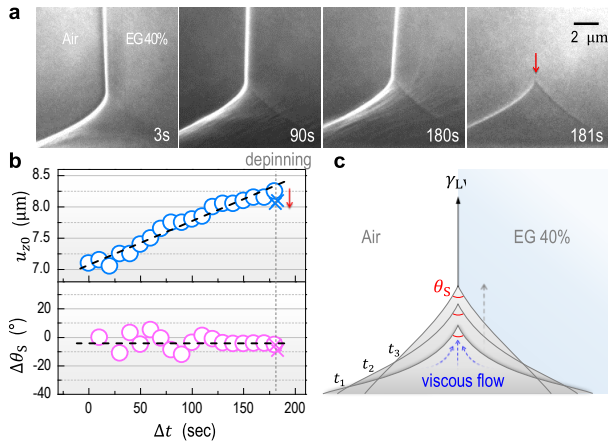

**Figure 3 | Ridge growth dynamics and its effect on cusp formation. a**, Representative sequential snapshots of a cusp during ridge growth for a EG 40% drop on a silicone gel ($E \approx 3$ kPa and $h \approx 50$ μm) (see Supplementary Video S1). The cusp was instantly recovered (red arrow) right after depinning at 181 s (gray dotted line). **b**, (Top) The ridge height $u_{z0}$ increases at a constant rate of ~7 nm/s until depinning at $\Delta t$ (observing time) = 181 s. The abrupt decrease right after depinning (red arrow) is attributed to an instant elastic recovery. (Bottom) $\Delta \theta_S$ is unchanged during the ridge growth ($\theta_S = 56.3 \pm 5.1°$). X-shaped symbols in **b** is the values obtained after depinning. Black dashed lines are a guide to the eye. **c**, A schematic illustration of ridge growth from $t = t_1$ to $t = t_3$. Invariant $\theta_S$ during the slow and linear ridge growth in **b** might be caused by a liquid-like viscous flow in the soft substrate.

**Ridge-growth dynamics.** To explore wetting dynamics on soft solids, ridge-growth dynamics should be first studied but is rarely known with few observations[16,17]. High temporal-resolution of TXM enabled us to take *real-time* movies of ridges during their growth. Figure 3a demonstrates a ridge growth for a nonvolatile EG 40% drop on silicone gel ($E \approx 3$ kPa), recorded 1~2 min after contact line pinning until the contact line is depinned by injecting EG 40% into the drop (see Supplementary Video S1). We plotted both the ridge height $u_{z0}$ and the change of the solid contact angle $\Delta \theta_S$ (within $w < ~0.7$ μm) as a function of observing time in Fig. 3b. We find that $u_{z0}$ linearly increases at a rate of ~7 nm/s. Here the abrupt decrease in $u_{z0}$ at 181 s (red arrows in Figs 3a and b) is due to the depinning. Very interestingly, the contact angles ($\theta_S$) at all the ridge tips observed were nearly invariant despite their gradual increase in $u_{z0}$, as demonstrated in Fig. 3b.

**Discussion**
Geometric invariance in the tip ($w \ll l_e$) during ridge growth or by surface softness indicates capillarity-controlled cusp formation[6,7]. For a large drop with negligible line tensions and Laplace pressures, only three interfacial tensions can affect ridge formation. We illustrated the force balances at the ridge tips for water and EG 40% drops in Fig. 4. Here we see that the formation of the asymmetric tips ($\theta_V \neq \theta_L$) is due to the asymmetric surface stresses ($\Upsilon_{SV} \neq \Upsilon_{SL}$). In fact, $\Upsilon_{SV}$ and $\Upsilon_{SL}$ can be significantly different although $\gamma_{SV} \approx \gamma_{SL}$ or $\theta \approx 90°$, as in Fig. 4b and Supplementary Table S2. The bending of the tips, clearly depicted in Fig. 4a, is attributed to the macroscopic force balance[8–10], as represented by $\theta$. These results suggest that the geometry of the cusp is determined by simultaneously applying macroscopic and microscopic force balances by Young's and Neumann's laws, respectively. Such a dual-scale force balance would be applicable to a general framework of wetting on soft solids within a wide softness range.

The time-invariant tip geometry, as in Fig. 3, implies that surface stresses are unchanged and thus the ridge growth observed would be an inelastic deformation[30]. The slow and linear growth of the ridges with the invariant contact angles ($\theta_S$) suggests that the ridge growth is, as illustrated in Fig. 3c, caused by a viscous flow in the soft solid, similar to that in a very viscous liquid[16].

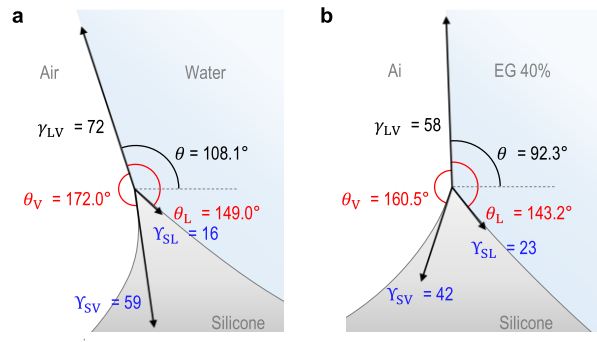

**Figure 4 | Microscopic force balances at the asymmetric and bent tips. a,b**, The estimated force balances at the asymmetric and bent tips for (**a**) water and (**b**) EG 40%, respectively. The liquid surface tension ($\gamma_{LV}$) and the surface stresses ($\Upsilon_{SL}$ and $\Upsilon_{SL}$) are given in mN/m.

In conclusion, we presented the first direct visualization of wetting ridge tips on soft solids using high-resolution x-ray microscopy (TXM). We revealed that the wetting ridges have asymmetric and bent tips. From the measurements of the microscopic and macroscopic contact angles at their tips, we found that the microscopic geometry of the tip is invariant during the ridge growth or by the surface elasticity. The singular geometry of the cusps can be deduced by simultaneously applying Young's and Neumann's laws that are valid, based on the surface stresses, for macroscopic and microscopic force balances, respectively. This report would open a new subject in wetting on soft solids, especially regarding study on ridge tip. Broadly speaking, x-ray microscopy would be helpful to investigate abnormal wetting behaviors on soft solids associated with spreading[14–16], contact angle hysteresis[17] or evaporation[17,18] and to study versatile elasto-capillary phenomena[19].

In this report, we suggested a simple approach of applying both Young's and Neumann's laws to dual-scale force balance, which would be potentially important, particularly, in natural systems. For example, cell-substrate interactions might be closely linked to wetting behaviors on soft solids[11]; in fact, cells are known to sense and respond to the environment through small strains (3~4%) that are induced by their traction forces[21]. Indeed, the mechanical microenvironment is, as seen in cancer progression in breast epithelial cells[20], critical to the normal development of most mammalian tissues and organs, which are soft viscoelastic materials ($E = 10^{-1}$ (e.g. a brain) ~ $10^2$ kPa (e.g. a cartilage)). In addition, our approach might be practically applicable to estimating the cell traction force or the force-induced strain[22] as well as the surface stresses[13].

**Methods**
The experiment was conducted using transmission x-ray microscopy (TXM; Fig. 1a)[31,32] at the 32-ID-C beamline in the Advanced Photon Source of the Argonne National Laboratory. We used a focused monochromatic x-ray beam at the photon energy of 9 keV (depth of focus: ~25 μm). The high brightness yielded high spatial-resolution (50 nm/pixel) in a short acquisition time (50 ms/frame). A Fresnel zone plate (ZP) was utilized as an objective lens. For phase contrast enhancement, we used a Au Zernike phase-ring (PR), optimized for 9 keV, which is very useful to imaging soft matters such as organic[33,34] or biological samples[31,32] without any contrast agents. Pure water (Millipore) or EG 40% (Samchun Pure Chemical Co., Korea) drops of $r \approx 1$ mm were generated by a syringe pump and gently put on a silicone gel ($E \approx 3$ kPa[6,13]; CY52-276A/B, Dow Corning Toray) or a PDMS ($E \approx 10$ or 16 kPa), spin coated with a controlled thickness ($h = 50$ μm) on a Si wafer (Fig. 1b). Si wafer as a solid support was used for the fine alignment of the samples to the x-ray beam, as confirmed by a mirror image of a wetting ridge (Fig. 2a) below the critical angle (0.22°) of Si. Image taking was started within 1~2 min after contact line pinning, under controlled temperature (22.5 °C) and humidity (19.5%). The contact angle $\theta$ for a rigid PDMS ($E \approx 750$ kPa) was, measured for 5 drops of water or EG 40% using optical microscopy, 106.6 ± 2.3° or 95.8 ± 0.7°, respectively. Effects of x-ray irradiation[35] to interfacial tensions or surface stresses are negligible, as confirmed by invariant contact angles.

**Acknowledgements**
We thank Eric R. Dufresne and Robert W. Style for helpful comments about this work. We also thank Guy K. German for preparing soft substrates. This research was financially supported by the Creative Research Initiatives (Functional X-ray Imaging) of MEST/NRF. Use of the Advanced Photon Source, an Office of Science User Facility operated for the U.S. Department of Energy (DOE) Office of Science by Argonne National Laboratory, was supported by the U.S. DOE under Contract No. DE-AC02-06CH11357.

# Supplementary information: Shedding new light on the mystery of wetting on soft solids

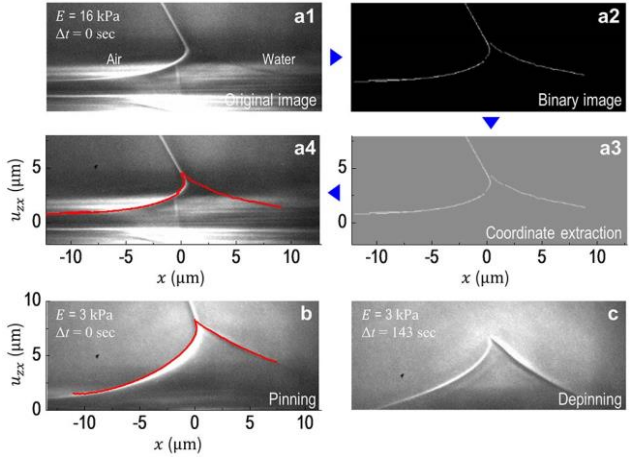

**Figure S1 | Extracting process of wetting ridge profiles. a1-4,** An extraction example of a ridge profile for $E \approx 16$ kPa. An original image (**a1**) was binarized, as shown in **a2**, by Canny edge detector in MATLAB. The coordinates of the white pixels in the binary image (**a2**) were then extracted, as in **a3**. The coordinates of the line points that could not be recognized by the Canny detector due to their weak contrasts were measured using a manual measurements tool in Image-Pro Plus 6.0 software. **a4,** The extracted ridge profile (red dots) is completely matched with the SL and SV interfaces in the original image (**a1**). **b,** The extracted profile of a ridge image for $E \approx 3$ kPa at $\Delta t = 0$ s. **c,** The ridge image obtained right after depinning the contact line in **b** ($\Delta t = 143$ s in this case). Compared with the tip profile in **b**, the clear tip image in **c** supports the validity of the extraction process. Here the interference bright and dark fringes at each interface in **a4** and **b** are originated from the Zernike phase contrast[31,32].

| Elasticity | No. | $\theta$ | $\theta_S$ | $\theta_V$ | $\theta_L$ |
|---|---|---|---|---|---|
| 3 kPa | 1 | 107.62 | 35.08 | 175.53 | 151.09 |
|  | 2 | 102.53 | 39.07 | 171.88 | 150.99 |
|  | 3 | 103.59 | 39.98 | 169.1 | 150.58 |
|  | 4 | 108.77 | 37.7 | 168.43 | 153.8 |
|  | 5 | 108.6 | 38.2 | 171.61 | 149.55 |
|  | Ave. | 106.22 | 38.01 | 171.31 | 151.2 |
|  | SD | 2.63 | 1.66 | 2.51 | 1.41 |
| 10 kPa | 1 | 108.91 | 37.45 | 177.12 | 145.29 |
|  | Ave. | 108.91 | 37.45 | 177.12 | 145.29 |
|  | SD | - | - | - | - |
| 16 kPa | 1 | 99.54 | 39.8 | 173.21 | 146.9 |
|  | 2 | 119.38 | 41.02 | 170.86 | 148.31 |
|  | 3 | 123.45 | 37.94 | 166.72 | 155.11 |
|  | Ave. | 114.12 | 39.59 | 170.26 | 150.11 |
|  | SD | 10.45 | 1.27 | 2.68 | 3.58 |

**Table S1 | Contact angles measured from x-ray images of wetting ridges on various soft substrates.** The macroscopic ($\theta$) and microscopic ($\theta_S$, $\theta_V$ and $\theta_L$) contact angles (°) were measured using Image-Pro Plus 6.0 software. For each elasticity condition, the average (Ave) and standard deviation (SD) were calculated.

| Liquids | $\gamma_{LV}$ | $\gamma_{SV}$ | $\gamma_{SL}$ | $\Upsilon_{SV}$ | $\Upsilon_{SL}$ |
|---|---|---|---|---|---|
| Water | 72 | 21* | 40* | 59 | 16 |
| EG 40% | 58 | 21* | 28† | 42 | 23 |

The properties were obtained from the literature (refs 36, 37).
*Ref. 36.
†Ref. 37.

**Table S2 | Interfacial tensions of liquids and estimated surface stresses (mN/m).**

**Video S1 | Wetting ridge growth for an EG 40% drop on a silicone gel ($E \approx 3$ kPa).** A wetting ridge shows a slow and linear growth tendency. See Figure 3 and text.